\documentclass[aps, prl, twocolumn, lengthcheck, superscriptaddress, 
nofootinbib]{revtex4-1}
\usepackage[utf8]{inputenc}
\usepackage{color}
\usepackage{rotating}
\usepackage{amsmath}
\usepackage{ulem}
\usepackage{overpic}
\usepackage{graphicx}
\usepackage{epstopdf}
\usepackage{dcolumn}
\usepackage{bm}
\usepackage{chngpage}
\usepackage{multirow}
\usepackage{slashed}
\usepackage{indentfirst}
\usepackage{booktabs}
\usepackage{amssymb,amsfonts}
\usepackage{amsmath}
\usepackage{color}
\usepackage{hyperref}
\usepackage{float}

\newcommand\sect[1]{{\it #1}---}

\begin{document}
\title{Doubly heavy tetraquark multiplets as heavy antiquark-diquark symmetry partners of heavy baryons}

\author{Tian-Wei Wu}
\email{wutianwei@ucas.ac.cn}
\affiliation{School of Fundamental Physics and Mathematical Sciences, Hangzhou Institute for Advanced Study, UCAS, Hangzhou, 310024, China}
\affiliation{University of Chinese Academy of Sciences, Beijing 100049, China}

\author{Yong-Liang Ma}
\email{ylma@ucas.ac.cn}
\affiliation{School of Fundamental Physics and Mathematical Sciences, Hangzhou Institute for Advanced Study, UCAS, Hangzhou, 310024, China}

\begin{abstract}
Symmetries play important roles in the understanding of hadron structures and spectroscopy. Motivated by the discovery of the doubly charmed tetraquark $T_{cc}^+(3875)$, we study the ground states of the doubly heavy tetraquarks with the QCD inspired heavy antiquark-diquark symmetry in the constituent quark model. Six ground states of $T_{QQ} (Q=c,b)$ are predicted and the lightest $T_{cc}$ state has a mass of $3875.8\pm 7.6$ MeV and spin-parity $1^+$ which are consistent with those of the observed $T_{cc}^+(3875)$. In addition, the magnetic moments of the predicted tetraquarks $T_{cc}^+(3876)$ and $T_{bb}^-(10396)$ are also estimated in the same model, which provide further informations to distinguish the structures of the $T_{QQ}$ states. Our results show that the heavy antiquark-diquark symmetry conserves well in these doubly heavy tetraquarks in both the spectrum and magnetic moment perspectives.
\end{abstract}

\maketitle

\sect{Introduction}
Since 2003, many exotic hadronic states which cannot be well understood by the conventional constituent quark model have been observed. Typically, in the heavy flavor sector, the observations of the so-called $X, Y, Z$ states, the $P_c$ pentaquarks and so on, have extended the hadron spectrums and deepen our understanding on the strong interactions between quarks in the nonperturbative region (see, e.g., Refs.~\cite{Hosaka:2016pey,Ali:2017jda,Guo:2017jvc,Esposito:2016noz,Lebed:2016hpi,Richard:2016eis,Chen:2016qju,Liu:2019zoy,Brambilla:2019esw} for reviews).
In 2017, the doubly heavy baryon $\Xi_{cc}^{++}$ was discovered in the
$\Lambda_c^+K^-\pi^+\pi^+$ mass spectrum by the LHCb collaboration with a mass $3621.4$ MeV~\cite{LHCb:2017iph}. Recently, the LHCb Collaboration reported a narrow state in the $D^0D^0\pi^+$ invariant mass spectrum~\cite{LHCb:2021vvq,LHCb:2021auc}, with $J^P=1^+$ and a mass $3875$~MeV very close to $D^{*+}D^0$  threshold, namely the $T_{cc}^+(3875)$. These first reported doubly heavy baryon and tetraquark arouse the interests of studies on the doubly heavy hadrons both theoretically and experimentally~~\cite{Ma:2015lba,Ma:2015cfa,Ma:2017nik,Shi:2017dto,Olamaei:2020bvw,Dong:2021bvy,Chen:2021vhg,Ren:2021dsi,Deng:2021gnb,Xin:2021wcr,Albaladejo:2021vln,Ling:2021bir,Meng:2021jnw,Fleming:2021wmk,Yan:2021wdl,Huang:2021urd,Du:2021zzh,Feijoo:2021ppq,Wu:2021kbu,Luo:2021ggs}.

Since in a doubly heavy hadron, the heavy quark is almost near its mass shell, it is natural to expect that the heavy quark limit is applicable. Therefore, the heavy diquark $\bar{X}=[QQ]$ in a doubly heavy hadron can be regarded as a compact object without radical excitation and belongs to the $\bar{3}_c$ color representation, the same as the one for the anti-heavy quark $\bar{Q}$.~\footnote{In this work, we change the notation $X\sim [QQ]$ used in \cite{Ma:2017nik} to $\bar{X}\sim [QQ]$ considering that $[QQ]$ is in the $\bar{3}_c$ representation.} Then, in the heavy quark limit, the color interactions are common for $\bar{X}$ and $\bar{Q}$, which leads to the superflavor symmetry, or the heavy antiquark-diquark symmetry (HADS), for the heavy quark sector~\cite{Savage:1990di,Georgi:1990ak,Carone:1990pv}.
The HADS sets up a relation between the hadrons with the same light quarks but different number of heavy quarks. There are two charm quarks in the doubly charmed baryon $\Xi_{cc}^{(*)}$, so that it can be related to $D^{(*)}$ meson through the HADS~\cite{Anselmino:1992vg,Cohen:2006jg,Ma:2017nik}.
With the HADS, one can easily derive a relation between the mass splitting of the doubly charmed baryon doublet and that of the charmed meson doublet, $m_{\Xi_{cc}^*}-m_{\Xi_{cc}}=\frac{3}{4} (m_{\bar{D}^*}-m_{\bar{D}})$~\cite{Hu:2005gf,Brambilla:2005yk,Ma:2017nik}, which has been numerically confirmed by a series of lattice QCD simulations~\cite{Padmanath:2015jea,Chen:2017kxr,Alexandrou:2017xwd,Mathur:2018rwu}. Again with the HADS, the systematic spectrum of the doubly heavy baryons is estimated by using the chiral partner structure and heavy quark spin-flavor symmetry~\cite{Ma:2017nik}.

In this work, we studied the spectrum of the doubly heavy tetraquarks using a constituent quark model with respect to the HADS in which the doubly heavy tetraquarks can be regarded as the HADS partners of the heavy baryons. Six ground tetraquark states with different quantum number configurations in both charm and bottom sectors are predicted. The predicted mass of the lowest doubly charmed tetraquark state, as a counterpart of $\bar{\Lambda}_{c}$ in the HADS, is consistent with the observed $T_{cc}^+(3875)$ . Although, whether the nature of $T_{cc}^+(3875)$ is a molecular state composed of $D$ and $D^*$ mesons or compact tetraquark is still on debate~\cite{Eichten:2017ffp,Karliner:2017qjm,Cheng:2020wxa,Weng:2021hje,Guo:2021yws,Chen:2022ros,Agaev:2021vur}. Since the HADS exists in the compact doubly heavy tetraquark picture but not in that of hadronic molecule, the doubly heavy tetraquarks predicted as the HADS partners of $\Sigma_{Q}^{(*)}$ can be easily distinguished form that predicted using the hadronic molecular picture.
As well as the mass relation between the doubly heavy baryons and heavy mesons mentioned above, HADS indicates specific relations between doubly heavy tetraquarks and heavy baryons. The observation of the predicted tetraquarks related to $\Sigma_{Q}^{(*)}$ baryons can be taken as an evidence of the tetraquark nature of $T_{cc}^+(3875)$.

Moreover, with the same model, the magnetic moments of the lowest doubly heavy tetraquarks are also estimated and compared to the calculation without the HADS. The results with and without HADS agree well, which indicates that the HADS conserves well in these doubly heavy systems. Along this line, considering that the observations of heavy baryons and doubly charmed tetraquark $T_{cc}^+(3875)$, it's very probable existing doubly heavy tetraquarks that are HADS partners of $\Sigma_{Q}^{(*)
}$ baryon.

\begin{figure}
    \centering
\includegraphics[width=8cm]{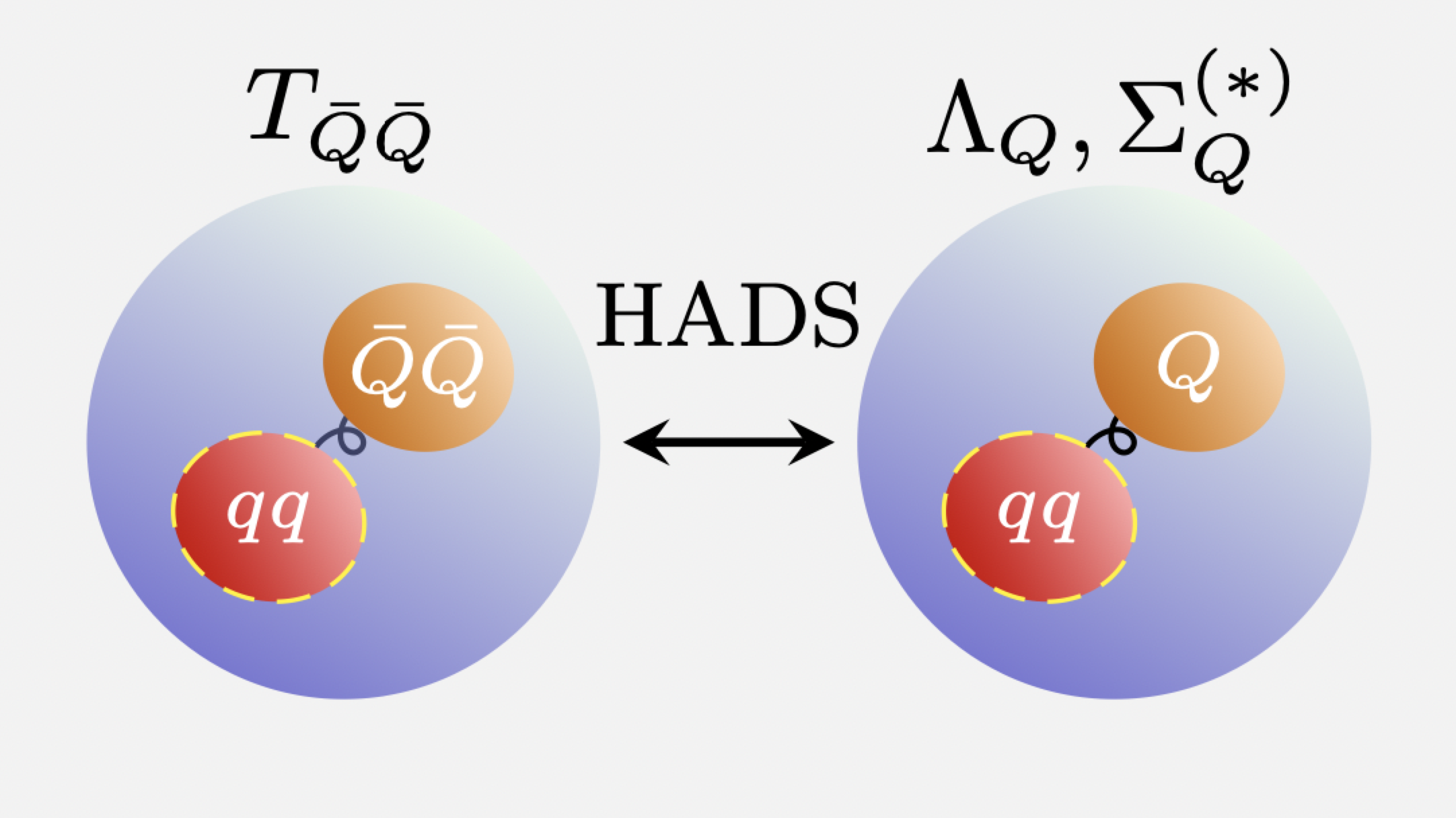}
    \caption{Cartoon of the HADS relations of $T_{\bar{Q}\bar{Q}}$ and $\Lambda_Q$, $\Sigma_Q^{(*)}$}
    \label{HADS}
\end{figure}

\sect{Model Description} In this work, we calculate the doubly heavy tetraquarks ($T_{cc}$ and $T_{bb}$) spectrum using the heavy antiquark-diquark symmetry (HADS) which relates the doubly heavy tetraquarks with the heavy antibaryons ($\bar{\Lambda}_{c}$, $\bar{\Sigma}_{c}^{(*)}$). According to the HADS~\cite{Savage:1990di,Georgi:1990ak,Carone:1990pv}, the heavy anti-diquark $[\bar{Q}\bar{Q}]$ can be regarded as a heavy object $X$ in $3_c$ representation. Along this line, there should exist doubly heavy tetraquarks $T_{[qq][\bar{Q}\bar{Q}]}$ ($q=u, d$ and $Q=c, b$) that can be regarded as the counterparts of $\Lambda_Q$ and $\Sigma_Q^{(*)}$ baryons, see Fig.~\ref{HADS}.

Therefore, thanks to the HADS, to obtained the spectrum of the doubly heavy tetraquarks, we only need to calculate the three-body system made of the heavy anti-diquark $X$ and the two light quarks instead of the system made of four quarks directly. The potential of this three-body system can be determined by using the constituent quark model for heavy baryon after substituting the heavy quark with $X$.

The mass formula of baryons in the constituent quark model reads
\begin{equation}
\label{MF}
    M=\sum_{i}m_i + \sum_{i<j}(F_i\cdot F_j)[B_{ij}+(\sigma_i\cdot\sigma_j)\alpha_{ij}/m_im_j],
\end{equation}
where $i,j=1,2,3$. Here $F_i$, $\sigma$ are the color and spin operators respectively.  The factor of color operator between two quarks in baryons is $-2/3$ ($\bar{3}_c$ representation). $B_{ij}$ is the binding between two quarks, while for light quark pair ($qq$) or heavy-light quark pair ($Qq$), the bindings are found to be 0~\cite{Karliner:2008sv,Karliner:2014gca}.  $m_i$ is the $i$-th quark mass and $\alpha$ is the quark-quark spin hyperfine interaction coupling. Note that the spin split of diquarks $\frac{\alpha}{m^2}$ decreases with quark masses ($m_q$, $m_s$, $m_c$, $m_b$), which is consistent with the convergence of heavy quark symmetry. The quark-quark spin hyperfine coupling $\alpha$ is the same for the light-light quarks and heavy-light quarks but is flavor dependent for the heavy diquark $QQ$. For detailed discussions, we refer to Refs.~\cite{Karliner:2008sv,Karliner:2014gca}.

In Eq.~\eqref{MF}, there are 7 parameters including four constituent quark masses($m_q$, $m_s$, $m_c$, $m_b$), two bindings($B_{cs}$, $B_{bs}$) and the coupling $\alpha$  to be determined from the ground baryon spectrum. We fix these parameters using a global best fit for the masses of all the ground baryons discovered so far (except $\Xi_{cc}$ which is taken as an input to constraint the binding between the two charm quarks). The global best fit strategy means that we determine the values of the parameters by minimizing the mass difference between the predicted and experimental values of all the 23 ground baryons. The parameters we obtained are listed
in Table~\ref{tab:Paras}. With these few parameters, the predicted masses in comparison with experimental values of ground baryon spectrum are shown in Table~\ref{tab:baryons}, with a smallest mass difference
\begin{equation}
   \chi_{\rm{Model}}=\sqrt{\sum_{i=1}^{23}(m_{\rm{pred}}-m_{\rm{exp}})^2/23}=7.6 \ \rm{MeV},
\end{equation}
which shows very a good consistence.

\begin{table}[t]
    \centering
    \caption{Masses, bindings and hyperfine couplings of quarks in baryons (in unit of MeV).}
    \begin{tabular}{c c c c c c c c c }
    \hline
    \hline
    Parameters& $m_q^b$& $m_s^b$& $m_c^b$ & $m_b^b$&$\alpha/(m_q^b)^2$&$B_{cs}$&$B_{bs}$\\
  Values&364.3&536.2&1715.9&5047.3& $-76.8$&$53.4$&$62.6$\\
    \hline
    \hline
    \end{tabular}
    \label{tab:Paras}
\end{table}

\begin{table*}[t]
    \centering
    \caption{Predicted masses of ground baryons with parameters in Table~\ref{tab:Paras} compared with experiment values (in unit of MeV).}
    \begin{tabular}{c c c c c}
    \hline
    \hline
    State& $I(J^+)$& $\sum_{i}m_i + \sum_{i<j}(F_i\cdot F_j)[B_{ij}+(\sigma_i\cdot\sigma_j)\alpha_{ij}/m_im_j]$& This model& PDG~\cite{ParticleDataGroup:2022pth} \\
    \hline
       N  &$1/2(1/2^+)$&$3m_q^b+2\alpha/(m_q^b)^2$ &939.3&938.9\\
    $\Delta$  &$1/2(3/2^+)$&$3m_q^b-2\alpha/(m_q^b)^2$ &1246.5&1232\\
    $\Lambda$  &$0(1/2^+)$&$2m_q^b+m_s^b+2\alpha/(m_q^b)^2$ &1111.2&1115.68\\
    $\Sigma$  &$1(1/2^+)$&$2m_q^b+m_s^b-\frac{2}{3}[\alpha/(m_q^b)^2-4\alpha/(m_q^bm_s^b)]$ &1176.9&1193.1\\
    $\Sigma^*$  &$1(3/2^+)$&$2m_q^b+m_s^b-\frac{2}{3}[\alpha/(m_q^b)^2+2\alpha/(m_q^bm_s^b)]$  &1385.6&1384.6\\
    $\Xi$  &$1/2(1/2^+)$&$2m_s^b+m_q^b-\frac{2}{3}[\alpha/(m_s^b)^2-4\alpha/(m_q^bm_s^b)]$  &1321.2&1318.3\\
    $\Xi^*$  &$1/2(3/2^+)$&$2m_s^b+m_q^b-\frac{2}{3}[\alpha/(m_q^s)^2+2\alpha/(m_q^bm_s^b)]$ &1529.9&1533.4\\
    $\Omega$  &$0(3/2^+)$&$3m_s-2\alpha/(m_{s}^2)$ &1679.5&1672.45\\
   $\Lambda_c$  &$0(1/2^+)$&$2m_q^b+m_c^b+2\alpha/(m_q^b)^2$ &2290.9&2286.46\\
    $\Sigma_c$  &$1(1/2^+)$&$2m_q^b+m_c^b-\frac{2}{3}[\alpha/(m_q^b)^2-4\alpha/(m_q^bm_c^b)]$ &2452.2&2453.54\\
      $\Sigma_c^*$  &$1(/2^+)$&$2m_q^b+m_c^b-\frac{2}{3}[\alpha/(m_q^b)^2+2\alpha/(m_q^bm_c^b)]$ &2517.4&2518.13\\
    $\Xi_{c}$  &$1/2(1/2^+)$&$m_c^b+m_q^b+m_s^b-\frac{2}{3}[B_{cs}-3\alpha/(m_s^bm_q^b)]$ &2476.4&2469.1\\
    $\Xi_{c}^{\prime}$  &$1/2(1/2^+)$&$m_c^b+m_q^b+m_s^b-\frac{2}{3}[B_{cs}+\alpha/(m_s^bm_q^b)-2\alpha/(m_c^bm_s^b)-2\alpha/(m_c^bm_q^b)]$ &2579.1&2578.5\\
    $\Xi_{c}^*$  &$1/2(3/2^+)$&$m_c^b+m_q^b+m_s^b-\frac{2}{3}[B_{cs}+\alpha/(m_s^bm_q^b)+\alpha/(m_c^bm_s^b)+\alpha/(m_c^bm_q^b)]$ &2633.8&2645.63\\

     $\Omega_{c}
    $  &$0(1/2^+)$&$m_c^b+2m_s^b-\frac{2}{3}[2B_{cs}+\alpha/(m_s^b)^2-4\alpha/(m_c^bm_s^b)]$ &2711.2&2695.2\\

      $\Omega_{c}^*
    $  &$0(3/2^+)$&$m_c^b+2m_s^b-\frac{2}{3}[2B_{cs}+\alpha/(m_s^b)^2+2\alpha/(m_c^bm_s^b)]$ &2755.5&2765.9\\

    $\Lambda_b$  &$0(1/2^+)$&$2m_q^b+m_b^b+2\alpha/(m_b^b)^2$ &5622.3&5619.6\\
    $\Sigma_b$  &$1(1/2^+)$&$2m_q^b+m_b^b-\frac{2}{3}[\alpha/(m_q^b)^2-4\alpha/(m_q^bm_b^b)]$ &5812.3&5813.1\\
    $\Sigma_b^*$  &$1(3/2^+)$&$2m_q^b+m_b^b-\frac{2}{3}[\alpha/(m_q^b)^2+2\alpha/(m_q^bm_b^b)]$ &5834.5&5832.53\\
        $\Xi_{b}$  &$1/2(1/2^+)$&$m_b^b+m_q^b+m_s^b-\frac{2}{3}[B_{bs}-3\alpha/(m_s^bm_q^b)]$ &5801.7&5794.5\\
    $\Xi_{b}^{\prime}$  &$1/2(1/2^+)$&$m_b^b+m_q^b+m_s^b-\frac{2}{3}[B_{bs}+\alpha/(m_s^bm_q^b)-2\alpha/(m_b^bm_s^b)-2\alpha/(m_b^bm_q^b)]$ &5928.4&5935.02\\
    $\Xi_{b}^*
    $  &$1/2(3/2^+)$&$m_b^b+m_q^b+m_s^b-\frac{2}{3}[B_{bs}+\alpha/(m_s^bm_q^b)+\alpha/(m_b^bm_s^b)+\alpha/(m_b^bm_q^b)]$ &5947.1&5955.32\\
      $\Omega_{b}
    $  &$0(1/2^+)$&$m_b^b+2m_s^b-\frac{2}{3}[2B_{bs}+\alpha/(m_s^b)^2-4\alpha/(m_b^bm_s^b)]$ &6049.8&6046.1\\
    \hline
    \hline
    \end{tabular}
    \label{tab:baryons}
\end{table*}

\sect{Masses of doubly heavy tetraquarks}
To study the mass spectrum of the doubly heavy tetraquarks,  we first denote their configurations as
\begin{equation}
\psi_T(C_{d},S_{d},C_{D},S_{D},I,J)=\left[[qq]^{C_d}_{S_d}[\bar{Q}\bar{Q}]^{C_D}_{S_D}\right]^I_{J},
\end{equation}
where $q=u,d$ is the light quark, $Q=c,b$ is the heavy quark, $C_d$($C_D$) and $S_d$($S_D$) are the color and spin quantum numbers of the light and heavy diquarks respectively.  $I$ is the total isospin and $J$ is the total angular momentum of the tetraquark.
Considering ground states of the doubly heavy tetraquarks, with the constraint of Pauli principle, we express the allowed configurations of $T_{[qq][\bar{Q}\bar{Q}]}$ as
\begin{equation}
\begin{split}
        \psi_{T[(0(1^+)]}&=\left[[qq]^{\bar{3}}_{0}[\bar{Q}\bar{Q}]^{3}_{1}\right]^0_{1},\\
        \psi_{T[(1(0^+)]}&=\left[[qq]^{\bar{3}}_{1}[\bar{Q}\bar{Q}]^{3}_{1}\right]^1_{0},\\
        \psi_{T[(1(1^+)]}&=\left[[qq]^{\bar{3}}_{1}[\bar{Q}\bar{Q}]^{3}_{1}\right]^1_{1},\\
        \psi_{T[(1(2^+)]}&=\left[[qq]^{\bar{3}}_{1}[\bar{Q}\bar{Q}]^{3}_{1}\right]^1_{2},\\
        \psi_{T'[(0(1^+)]}&=\left[[qq]^{6}_{1}[\bar{Q}\bar{Q}]^{\bar{6}}_{0}\right]^0_{1},\\
        \psi_{T'[(1(0^+)]}&=\left[[qq]^{6}_{0}[\bar{Q}\bar{Q}]^{\bar{6}}_{0}\right]^1_{0}.\\
\end{split}
\end{equation}

In these six configurations, if we regard the heavy antiquark pair $X=[\bar{Q}\bar{Q}]$ as a compact color source, in the sense of the HADS, the first tetraquark with isospin 0 and spin-parity $1^+$ should be the counterpart of $\Lambda_Q$ and the second to the fourth ones with isospin 1 and spin-parity $0^+, 1^+, 2^+ $ are the counterparts of $\Sigma_Q^{(*)}$, because both these heavy tetraquarks and baryons have the same quantum number configurations of the light diquark.  While for the last two tetraquarks, they are in the novel color structures that do not appear in mesons and baryons because of the color confinement, which could be good subjects to study the interactions and properties in new color structures.

To calculate the mass of $T_{[qq][\bar{Q}\bar{Q}]}$ with mass formula in Eq.~(\ref{MF}) by regarding the heavy anti-diquark $X=[\bar{Q}\bar{Q}]$ as a heavy boson-like quark, we need two more parameters $\alpha_{QQ}$ and $B_{QQ}$, and the coefficients of color and spin operators. The $\alpha_{cc}$ can be determined by the masses of charmonium $J/\psi$ and $\eta_c$ as $\alpha_{cc}/m_c^2={}-3/16(m_{J/\psi}-m_{\eta_{c}})={}-21.2$ MeV. And $B_{cc}$ can be estimated through the mass formula $m_{\Xi_{cc}}=2m_c^b+m_q^b-2/3[B_{cc}+\alpha_{cc}/(m_q^b)^2-4\alpha/{(m_q^bm_c^b)}]$ as $217.7$~MeV. Similarly, the parameter $\alpha_{bb}$ can be determined as $\alpha_{bb}/m_b^2=-3/16(m_{\Upsilon_b}-m_{\eta_{b}})=-11.6$ MeV. While for $B_{bb}$, since there is no experimental data of doubly bottomed baryon,  we instead use the heavy bottomonium data, $B_{bb}=-3/16[(3m_{\Upsilon_b}+m_{\eta_{b}})-8m_b^m]=422.0$ MeV~\cite{Karliner:2014gca}. With the parameters and mass formula in Eq.~(\ref{MF}), the masses of the  doubly heavy baryons are predicted to be $\Xi_{cc}^*=3686.8$ MeV, $\Xi_{bb}=10170.5$ MeV and $\Xi_{bb}^*=10192.6$ MeV.
The color operator coefficients in Eq.~\eqref{MF} are trial, and the spin operator coefficients serves
$\sum_{i<j}\sigma_i\cdot\sigma_j=4\sum_{i<j}S_i\cdot S_j=2(S^2-\sum_{j}S_j^2)$.
With
$\sigma_{q_1}\cdot\sigma_{q_2}$ and $\sigma_{\bar{Q}_3}\cdot\sigma_{\bar{Q}_4}$ clarified, $\sigma_{q}\cdot\sigma_{\bar{Q}\bar{Q}}$ can be calculated by $(\sum\sigma_i\cdot\sigma_j-\sigma_{q_1}\cdot\sigma_{q_2}-\sigma_{\bar{Q}_3}\cdot\sigma_{\bar{Q}_4})/2$.

For the heavy diquark $\bar{X}=[QQ]^C_S$ that can be regarded as a heavy antiquark in the sense of HADS, its mass is expressed as
\begin{equation*}
    m_{[QQ]^C_S}=2m_Q+(F_1\cdot F_2)[B_{QQ}+(\sigma_1\cdot \sigma_2)\alpha_{QQ}/m_Q^2].
\end{equation*}
With the parameters in Table~\ref{tab:Paras}, we have $m_{[cc]^{\bar{3}}_1}=3300.8$~MeV, $m_{[cc]^{6}_0}=3525.6$~MeV,
$m_{[bb]^{\bar{3}}_1}=9821.0$~MeV and
$m_{[bb]^{6}_0}=10246.8$~MeV. For the coupling between heavy diquarks $[QQ]^C_S$ and light quark $q$, according to HADS in heavy quark limit, it's the same as $\alpha_{Qq}$, so we have $\alpha_{[QQ]^C_Sq}=\alpha$.

 Once the masses of heavy diquarks and the coupling are determined, we can use Eq.~(\ref{MF}) to calculate the mass spectrum of the doubly charmed tetraquarks $T_{cc}$ and the doubly bottomed tetraquarks $T_{bb}$. The  predicted masses and their expressions of different quantum number configurations are summarized in Table~\ref{tab: Mass}.

\begin{table*}[t]
    \centering
    \caption{Predicted masses (in unit of MeV) of doubly heavy tetraquarks.}
    \begin{tabular}{c c c c c c}
    \hline
    \hline
    State& $I(J^P)$&Configuration & Mass formula& $T_{cc}$& $T_{bb}$\\
    \hline
    $T_{QQ}$  &0$(1^+)$&$\left[[\bar{q}\bar{q}]^{3}_{0}[QQ]^{\bar{3}}_{1}\right]^0_{1}$ &$2m_q^b+m_{[QQ]^{\bar{3}}_1}+2\alpha/(m_q^b)^2$ &$3875.8\pm 7.6$& $10396.0\pm 7.6$\\
    $T_{QQ}$  &$1(0^+)$&$\left[[\bar{q}\bar{q}]^{3}_{1}[QQ]^{\bar{3}}_{1}\right]^1_{0}$&$2m_q^b+m_{[QQ]^{\bar{3}}_1}-\frac{2}{3}[\alpha/(m_q^b)^2-8\alpha_{[QQ]^{\bar{3}}_1q}/(m_q^bm_{[QQ]^{\bar{3}}_1})]$ &$4035.4\pm 13.6$&$10585.6\pm 8.5$\\
    $T_{QQ}$  &$1(1^+)$&$\left[[\bar{q}\bar{q}]^{3}_{1}[QQ]^{\bar{3}}_{1}\right]^1_{1}$&$2m_q^b+m_{[QQ]^{\bar{3}}_1}-\frac{2}{3}[\alpha/(m_q^b)^2-4\alpha_{[QQ]^{\bar{3}}_1q}/(m_q^bm_{[QQ]^{\bar{3}}_1})]$ &$4058.0\pm 9.5$&$10593.2\pm 7.8$\\
    $T_{QQ}$  &$1(2^+)$&$\left[[\bar{q}\bar{q}]^{3}_{1}[QQ]^{\bar{3}}_{1}\right]^1_{2}$&$2m_q^b+m_{[QQ]^{\bar{3}}_1}-\frac{2}{3}[\alpha/(m_q^b)^2+4\alpha_{[QQ]^{\bar{3}}_1q}/(m_q^bm_{[QQ]^{\bar{3}}_1})]$ &$4103.2\pm 9.5$&$10608.3\pm 7.8$\\
    $T_{QQ}'$  &$0(1^+)$&$\left[[\bar{q}\bar{q}]^{{\bar{6}}}_{1}[QQ]^{{6}}_{0}\right]^0_{1}$&$2m_q^b+m_{[QQ]^6_0}+\frac{1}{3}\alpha/(m_q^b)^2$&$4228.6\pm 7.6$&$10949.8\pm 7.6$\\
    $T_{QQ}'$  &$1(0^+)$&$\left[[\bar{q}\bar{q}]^{{\bar{6}}}_{0}[QQ]^{{6}}_{0}\right]^1_{0}$&$2m_q^b+m_{[QQ]^6_0}-\alpha/(m_q^b)^2$&$4331.0\pm 7.6$&$11052.2\pm 7.6$\\
    \hline
    \hline
    \end{tabular}
    \label{tab: Mass}
\end{table*}

The lowest mass of the predicted doubly charmed tetraqurark $T_{cc}$ is found to be $3875.8\pm7.6$ MeV, which is consistent with the discovered mass of $T_{cc}^+(3875)$~\cite{LHCb:2021vvq,LHCb:2021auc}.
The heavy quark spin symmetry indicates that the splits of the spin multiplets of the doubly heavy tetraquarks converge when the heavy quark limit is taken. The predicted mass spectrum also shows this convergence---the mass splitings between $T_{bb}$ isovector multiplets is about one third of that between $T_{cc}$ isovector multiplets, which is actually the mass ratio of constituent charm and bottom quarks ($\frac{m_b}{m_c}=2.94$).
Same as the relation between the mass splitting of doubly charmed baryons and that of the charmed mesons, i.e., $m_{\Xi_{cc}^*}-m_{\Xi_{cc}}=\frac{3}{4} (m_{\bar{D}^*}-m_{\bar{D}})$~\cite{Hu:2005gf,Brambilla:2005yk,Ma:2017nik}, one can deduce a relation between the mass splitting of the isovectoral heavy baryons and that of the doubly heavy tetraquarks
\begin{equation}
    m_{T_{QQ}[1(2^+)]}-m_{T_{QQ}[1(0^+)]}
    =m_{\Sigma_Q^*}-m_{\Sigma_Q},
\end{equation}
which is exactly the case of the predictions of this model.
Moreover, one can also deduce the mass relation between the isovectoral and isoscalar states of the ground doubly heavy tetraquarks and the ground heavy baryons as
\begin{eqnarray}
& & 2m_{T_{QQ}[(1(2^+)]}+m_{T_{QQ}[1(0^+)]}-3m_{T_{QQ}[0(1^+)]}\nonumber\\
& & \qquad\qquad\qquad\qquad\qquad = 2m_{\Sigma_c^*}+m_{\Sigma_c}-3m_{\Lambda_c}.
\end{eqnarray}
 These mass relations are based on the HADS and not shared with the molecular picture. Therefore they can not only be used to clarify their nature, but also be evidence to verify or defuse the molecular nature of the observed $T_{cc}^+(3875)$, in case that these predicted doubly heavy tetraquarks are discovered in future experiments.

We next consider the uncertainties of the calculation which essentially come from two aspects, the uncertainty of the model and the breaking of HADS. The uncertainty of the model can be estimated by the average mass difference $\chi_{\rm{Model}}$ between the predictions and experimental values presented in Table~\ref{tab:baryons}, $\sim 7.6$~MeV.
The breaking of HADS is at the level of $\Lambda_{QCD}/(m_Q v)$~\cite{Savage:1990di}, where $v$ is the velocity of the heavy quark pair. Following Refs.~\cite{Hu:2005gf,Wu:2020rdg}, we consider a 25\% breaking of HADS. The breaking of HADS only affects the hyperfine coupling $\alpha_{[QQ]^{C}_{S}q}$, thus the uncertainty of HADS breaking can be estimated by
\begin{equation}
   \chi_{\rm{HADS}}=0.25 (F_{[QQ]^{C}_{S}}\cdot F_q) (\sigma_{[QQ]^{C}_{S}}\cdot \sigma_q)\frac{\alpha_{[QQ]^{C}_{S}q}}{m_q^bm_{[QQ]^{C}_{S}}},
\end{equation}
which yields $\sim 11.3$~MeV and $\sim3.8$~MeV for $T_{cc}$ and $T_{bb}$ in $1(0^+)$ state and the half of that in $1(1^+)$ and $1(2^+)$ states, respectively. Assuming these two uncertainties are independent, the total uncertainty is then estimated by  $\chi= \sqrt{\chi_{\rm{Model}}^2+\chi^2_{\rm{HADS}}}$.

Another interesting thing is that the HADS is naturally conserved in the present calculation for $T_{QQ}[0(1^+)]$, $T_{QQ}^{\prime}[0(1^+)]$ and $T_{QQ}^{\prime}[1(0^+)]$ states, because the HADS affected term $\alpha_{[QQ]^{C}_{S}q}/{m_Q^2}$ does not appear in their mass expressions. This means that these states can be directly predicted from the mass formula in Eq.~(\ref{MF}) by extending it to four body.

\sect{Magnetic moment}
The magnetic moment of a hadron is expressed as
\begin{equation}
\label{mm}
    \mu_{H}=\sum_{i}\langle H\uparrow|2\mu_i s_{zi|}H \uparrow\rangle,
\end{equation}
where $\mu_i=g\frac{q_ie}{2m_i}s$ is the magnetic moment of quark $i$ with charge $q_ie$ and $g=2$ for a point particle. $s_{zi}$ is the $z$-axis component of the spin operator. $|H \uparrow\rangle$ is the flavor spin wave function of the hadron. With expression~\eqref{mm},
the magnetic moment of $\Lambda_Q$ is then predicted to be $\mu_{\Lambda_Q} =\mu_Q=\frac{q_
Qe}{2m_Q}$. With HADS, the heavy diquark $\bar{X}=[QQ]^{\bar{3}}_1$ can be viewed as a heavy antiquark with color $\bar{3}$ and spin 1. The same as the $\Lambda_Q$, the magnetic moment of the predicted $0(1^+)$ doubly tetraquark $T_{[[\bar{u}{\bar{d}}]^{3}_0[QQ]^{\bar{3}}_1]}$ can be estimated as $\mu_{T_{[QQ]^{\bar{3}}_1}}=\mu_{[QQ]^{\bar{3}}_1}=\frac{q_
{[QQ]^{\bar{3}}_1}e}{m_{[QQ]^{\bar{3}}_1}}$. With the mass of $[QQ]^{\bar{3}}_1$ obtained above, the magnetic moments of $T_{cc}^+(3876)$ and $T_{bb}^-(10396)$ are
\begin{equation*}
\label{mmHADS}
    \mu_{T_{cc}^+(3876)}=0.759\mu_N,\quad \mu_{T_{bb}^-(10396)}=-0.127\mu_N,
\end{equation*}
which are consistent with the light-cone QCD sum rule calculation of the $J^P=1^+$ tetraquark state using the diquark-antidiquark picture, $\mu_{T_{cc}^+-\rm{Di}}=0.66^{+0.34}_{-0.23}$~\cite{Azizi:2021aib}.

As the diquark $[QQ]^{\bar{3}}_1$ is not a genuine point-like quark, we can also use Eq.~(\ref{mm}) to calculate the magnetic moments of the lowest $T_{cc}^+$ and $T_{bb}^-$ with quantum numbers $0(1^+)$. The flavor-spin wave function reads
\begin{eqnarray}
    |T_{[[\bar{u}{\bar{d}}]^{3}_0[QQ]^{\bar{3}}_1]^0_1}\uparrow\rangle & = & \left(\frac{1}{\sqrt{2}}|\bar{u}\bar{d}-\bar{d}\bar{u} \rangle\frac{1}{\sqrt{2}}|\uparrow\downarrow-\downarrow\uparrow\rangle \right)\nonumber\\
    & & \qquad \times |QQ\uparrow\uparrow\rangle.
\end{eqnarray}
The magnetic moment of $T_{[[\bar{u}{\bar{d}}]^{3}_0[QQ]^{\bar{3}}_1]^0_1}$ is therefore obtained as $2\mu_{Q}=\frac{q_Qe}{m_Q}$. With the mass of heavy quark $m_Q$ in Table~\ref{tab:Paras}, the magnetic moments of $T_{cc}^+(3876)$ and $T_{bb}^-(10396)$ are
\begin{equation*}
\label{mmTQQ}
    \mu_{T_{cc}^+(3876)}=0.732\mu_N,\quad \mu_{T_{bb}^-(10396)}=-0.124\mu_N,
\end{equation*}
which are very close to the HADS prediction and implies that the HADS also conserves in magnetic moment perspective.

\sect{Conclusions and discussions} In this paper, we construct a simple but effective model based on the QCD inspired heavy antiquark-diquark symmetry(HADS) to study the doubly heavy tetraquarks. Thanks to the HADS,  the doubly heavy tetraquarks are related to the heavy baryons, of which the masses are predicted with an explicit mass formula. Six ground states of both doubly charmed and bottomed tetraquarks are predicted.

The results show that the predicted mass of the lowest doubly charmed tetraquarks $T_{cc}(3876)$ is well consistent with the observed $T_{cc}^+(3875)$ by LHCb. The isoscalar $T_{QQ}[0(1^+)]$ can be viewed as the HADS partner of $\bar{\Lambda}_Q$, while the isovectoral multiplets $T_{QQ}[1(0^+)], T_{QQ}[1(1^+)]$ and $T_{QQ}[1(2^+)]$ are the HADS partners of $\bar{\Sigma}_{Q}^{(*)}$, for which we propose a high possibility that they could exist. For the $T_{QQ}^{\prime}[0(1^+)]$ and $T_{cc}^{\prime}[1(0^+)]$, they are in novel color structures that do not occur in mesons and baryons, which may provide further understandings about the strong interactions between quarks if discovered experimentally.

Within the HADS, the mass relations between the doubly heavy tetraquarks and the heavy baryons are also studied, which can be tested by future experiments.
Due to heavy quark flavor symmetry, these studies are also suitable for doubly heavy tetraquarks that contain both charm and bottom quarks.
Although we do not discuss the doubly heavy tetraquarks $T_{bc}$ in this work because of the absence of doubly heavy baryons $\Xi_{bc}^{(*)}$ and the lack of bottom-charmed mesons, it's easy to follow these studies when the poor situations of data between charm and bottom quarks have changed.

By using the same model, we also estimated the magnetic moments of the isoscalar tetraquarks $T_{cc}^+(3876)$ and $T_{bb}^-(10396)$ states and compared to the calculation without the HADS. The results show good consistence, which indicates that the HADS conserves well in these doubly heavy tetraquarks. The predicted magnetic moments are also consistent with other previous studies in diquark-antidiquark picture but differ from the hadronic molecular picture. Thus the magnetic moment provides further informations to distinguish the nature of $T_{cc}^+(3875)$.
Along this line, considering that the observations of heavy baryons and doubly charmed tetraquark $T_{cc}^+(3875)$, it's very probable existing doubly heavy tetraquark multiplets that are HADS partners of $\Sigma_{Q}^{(*)}$ baryon. We encourage our colleagues to search for these states in the current and upcoming facilities.

{\it Acknowledgements}: T. W. W is supported by the National Natural Science Foundation of China under Grants No. 12147152 and China Postdoctoral
Science Foundation under Grants No. 2022M723119. Y. L. M. is supported in part by the National Science Foundation
of China (NSFC) under Grants No. 11875147 and No.
12147103.

\bibliography{doubly-heavy}

\end{document}